\documentclass[aps,prmaterials,superscriptaddress,reprint,amsmath,amssymb,longbibliography,floatfix]{revtex4-2}

\usepackage[utf8]{inputenc}
\usepackage[T1]{fontenc}
\usepackage{amsmath}
\usepackage{amssymb}
\usepackage{graphicx}
\usepackage{epstopdf}
\usepackage{color}
\usepackage{enumerate}
\usepackage{ulem}
\usepackage{multirow}
\usepackage{array}
\usepackage{float}
\usepackage{amsmath}
\usepackage{hyperref}
\usepackage{epsfig}
\usepackage{bm}
\usepackage{hyphenat}
\usepackage{makecell}
\usepackage{color,soul}
\usepackage[dvipsnames]{xcolor}
\usepackage{chemformula}
\usepackage{siunitx}
\usepackage{xcolor}
\usepackage{comment}
\DeclareSIUnit\angstrom{\text {Å}}

\begin{document}

\title{Fermi-surface studies of altermagnetic CrSb from Shubnikov-de Haas oscillations}

\author{Sajal Naduvile Thadathil}
\affiliation{Hochfeld-Magnetlabor Dresden (HLD-EMFL) and W\"urzburg-Dresden
Cluster of Excellence ctd.qmat, Helmholtz-Zentrum
Dresden-Rossendorf, 01328 Dresden, Germany}
\affiliation{Institut f\"ur Festk\"orper- und Materialphysik,
Technische Universit\"at Dresden, 01062 Dresden, Germany}

\author{Beat Valentin Schwarze}
\affiliation{Hochfeld-Magnetlabor Dresden (HLD-EMFL) and W\"urzburg-Dresden
Cluster of Excellence ctd.qmat, Helmholtz-Zentrum
Dresden-Rossendorf, 01328 Dresden, Germany}

\author{Jaafar Ansari }
\affiliation{Max Planck Institute for the Physics of Complex Systems, Nöthnitzer Str. 38, 01187 Dresden, Germany}

\author{Tommy Kotte}
\affiliation{Hochfeld-Magnetlabor Dresden (HLD-EMFL) and W\"urzburg-Dresden
Cluster of Excellence ctd.qmat, Helmholtz-Zentrum
Dresden-Rossendorf, 01328 Dresden, Germany}

\author{Sven Luther}
\affiliation{Hochfeld-Magnetlabor Dresden (HLD-EMFL) and W\"urzburg-Dresden
Cluster of Excellence ctd.qmat, Helmholtz-Zentrum
Dresden-Rossendorf, 01328 Dresden, Germany}

\author{Marc Uhlarz}
\affiliation{Hochfeld-Magnetlabor Dresden (HLD-EMFL) and W\"urzburg-Dresden
Cluster of Excellence ctd.qmat, Helmholtz-Zentrum
Dresden-Rossendorf, 01328 Dresden, Germany}

\author{Freya Husstedt}
\affiliation{Hochfeld-Magnetlabor Dresden (HLD-EMFL) and W\"urzburg-Dresden
Cluster of Excellence ctd.qmat, Helmholtz-Zentrum
Dresden-Rossendorf, 01328 Dresden, Germany}
\affiliation{Institut f\"ur Festk\"orper- und Materialphysik,
Technische Universit\"at Dresden, 01062 Dresden, Germany}

\author{Rafael Gonzalez-Hernandez}
\affiliation{Departamento de Fisica y Geociencias, Universidad del Norte, Barranquilla, Colombia}

\author{Libor \v{S}mejkal}
\affiliation{Max Planck Institute for the Physics of Complex Systems, Nöthnitzer Str. 38, 01187 Dresden, Germany}
\affiliation{Max Planck Institute for Chemical Physics of Solids, Nöthnitzer Str. 40, 01187 Dresden, Germany}
\affiliation{Institute of Physics, Czech Academy of Sciences, Cukrovarnická 10, 162 00 Prague 6, Czech Republic}

\author{Thanassis Speliotis}
\affiliation{Institute of Nanoscience and Nanotechnology, National Center for Scientific Research Demokritos, 15341 Athens, Greece}

\author{Markéta Žáčková}
\affiliation{Faculty of Mathematics and Physics, Charles University, 12116 Prague, Czech Republic}

\author{Jiří Pospíšil}
\affiliation{Faculty of Mathematics and Physics, Charles University, 12116 Prague, Czech Republic}

\author{Christoph M\"uller}
\affiliation{Institute of Physics, Czech Academy of Sciences, Cukrovarnická 10, 162 00 Prague 6, Czech Republic}
\affiliation{Faculty of Mathematics and Physics, Charles University, 12116 Prague, Czech Republic}

\author{Dominik Kriegner}
\affiliation{Institute of Physics, Czech Academy of Sciences, Cukrovarnická 10, 162 00 Prague 6, Czech Republic}

\author{Helena Reichlov\'a}
\affiliation{Institute of Physics, Czech Academy of Sciences, Cukrovarnická 10, 162 00 Prague 6, Czech Republic}

\author{Jochen Wosnitza}
\affiliation{Hochfeld-Magnetlabor Dresden (HLD-EMFL) and W\"urzburg-Dresden
Cluster of Excellence ctd.qmat, Helmholtz-Zentrum
Dresden-Rossendorf, 01328 Dresden, Germany}
\affiliation{Institut f\"ur Festk\"orper- und Materialphysik,
Technische Universit\"at Dresden, 01062 Dresden, Germany}

\author{Toni Helm}
\email{t.helm@hzdr.de}
\affiliation{Hochfeld-Magnetlabor Dresden (HLD-EMFL) and W\"urzburg-Dresden
Cluster of Excellence ctd.qmat, Helmholtz-Zentrum
Dresden-Rossendorf, 01328 Dresden, Germany}

\date{\today}

\begin{abstract}

Within the family of altermagnets, CrSb is a metallic, collinearly ordered material that exhibits particularly strong symmetry-induced spin splitting in its band structure.
In this study, we combine electrical magnetotransport measurements up to 68\,T on microfabricated single-crystalline CrSb with first-principles calculations to investigate its Fermi surface.
Notably, we study the temperature and field-orientation dependence of magnetic quantum oscillations observed in the magnetoresistance.
The observed frequency spectrum agrees well with results from density-functional-theory calculations that take spin-orbit coupling into account, without invoking significant ad hoc band shifts.
Our results confirm the predicted electronic band structure of altermagnetic CrSb 
and highlight the importance of high magnetic fields for accurately mapping the Fermi surfaces of unconventional emergent materials.

\end{abstract}

\maketitle
\section{Introduction}
   
Altermagnets (AMs) have recently been predicted as a third fundamental class of magnetic materials, distinct from conventional ferromagnets and antiferromagnets~\cite{Smejkal2022,Smejkal2022b,Mazin2022}.
In these materials, a compensated collinear magnetic structure coexists with a large, momentum‑dependent spin splitting of the electronic bands that are exchange driven and can exist even in the absence of spin-orbit coupling (SOC)~\cite{krempasky2024,lee2024,osumi2024,reimers2024,fedchenko2024, Jiang2025}.
This nonrelativistic spin splitting is enabled by unconventional spin–space group symmetries, which lift the Kramers degeneracy without relying on strong spin–orbit coupling~\cite{krempasky2024}.
Theoretical studies further predict that many AMs exhibit topological semimetal phases, including Weyl nodes, nodal lines, and higher‑order topological states~\cite{Li2025,Lu2025, jungwirth2024, Shuai2025, Daniil2025, Li2024}. 
Consequently, altermagnets provide a promising platform, in which magnetic order, symmetry, and band topology are intertwined in ways not achievable in conventional magnetic or nonmagnetic materials~\cite{Liu2025, Liu2025chern}.

Several compounds have now been experimentally confirmed to host the AM state~\cite{Smejkal2022, Guo2023}. Well recognized are, for example, MnTe~\cite{lee2024,osumi2024,aoyama2024,krempasky2024,gonzalez2023}, CrSb~\cite{reimers2024, urata2024, Bommanaboyena2025}, and FeS~\cite{Takagi2025}.
Among these, CrSb stands out due to its high spin-splitting energy (1.2\,eV)~\cite{Smejkal2022,Guo2023} and high N\'eel temperature (700\,K)~\cite{Takei1963,Snow1952}, rendering it a room-temperature altermagnet.
CrSb crystallizes in the hexagonal NiAs‑type structure (P6$_3$/mmc space group).
It is identified as a prototypical $g$‑wave altermagnet, as confirmed by ARPES measurements~\cite{reimers2024, Yang2025}.
Moreover, experimental studies have revealed pronounced spin-split bands and spin-polarized arc-like surface states in CrSb~\cite{Lu2025, Li2025}.
These observations point to an altermagnetic Weyl-semimetallic state with substantial exchange‑driven band splitting near the Fermi level. 
These results position CrSb as a potential platform to probe how altermagnetic symmetry shapes bulk and surface electronic structures.

Magnetotransport measurements provide a sensitive bulk probe of such topological and multiband electronic structures~\cite{Leahy2018, Shekhar2015, Mizuku2021}.
In many semimetals, extremely large, nonsaturating magnetoresistance and nonlinear Hall response have been linked to compensated multicarrier transport and nontrivial band topology.
Similar behavior has recently been reported in CrSb~\cite{urata2024,bai2025,peng2025}.
Studies at low magnetic fields revealed that the Hall resistivity is strongly nonlinear, indicating the presence of multiple charge-carrier pockets with distinct mobilities.
Also, a nonsaturating magnetoresistance is reported~\cite{urata2024,bai2025,peng2025}. 

High-magnetic-field measurements further extend this picture by directly mapping the extremal Fermi‑surface (FS) cross sections via an analysis of magnetic quantum oscillations (MQOs)~\cite{Shoenberg1984}.
Recent measurements on CrSb up to 35\,T and field aligned along the $c$ direction report multiple Shubnikov-de Haas (SdH) frequencies, indicating several distinct FS pockets with light effective masses of the order of about one free electron mass~\cite{Du2025}.
In addition, nontrivial Berry phases for several orbits and Zeeman‑driven Landau‑level splitting were reported.
In two other MQO studies up to 41.5\,T, including full angular rotations through nodal/antinodal planes, the bulk $g$-wave spin-split FS sheets have been mapped~\cite{Long2026, Terashima2026}.
The works argue to have confirmed the alternating symmetry of the magnetic order-parameter via frequency-split patterns~\cite{Long2026}. 
Yet, the detailed evolution of these SdH oscillations with field and temperature remains only partially understood.

Given its complex electronic, magnetic, and topological properties, high-field magnetotransport measurements are crucial for studying CrSb, since high pulsed fields allows to probe fundamental transport behavior well beyond the limits of conventional steady-field experiments.
Therefore, the present study focuses on high‑field electrical magnetotransport and magnetic quantum oscillations in single‑crystalline microstructured CrSb.
By combining longitudinal- and Hall-resistivity measurements in steady and pulsed fields with a detailed SdH analysis, it becomes possible to resolve the underlying FS, extract effective masses, and mobilities.
Placing these quantum‑oscillation results alongside recent ARPES and band‑structure calculations allows to gain a more complete picture of the FS of CrSb.
It should be noted that ARPES is a surface-sensitive technique and bulk probes, such as MQO measurements, are highly desirable to determine the bulk Fermi surface.

\section{Experimental Methods}

We synthesized high-quality single crystals of CrSb by chemical vapor transport.
The crystals originated from the same growth batch as those employed in previous XMCD investigation~\cite{Biniskos2025}.

We fabricated CrSb microstructures using gallium focused-ion-beam (FIB) lithography.
First, we cut out a lamella-shaped piece with dimensions $(32 \times 2.4 \times 130)\,\mu\mathrm{m}^3$ from the bulk crystal using a Ga-FIB system and transferred it ex-situ onto a sapphire substrate.
Next, we deposited an approximately \SI{150}{\nano\meter}-thick layer of gold over the entire lamella by magnetron sputtering.
After that, the gold layer was partially etched away from the central top surface of the sample by Ga$^{+}$ ions.
Then, a FIB was used to cut trenches into the gold layer and the sample to create well-defined contact terminals.
We achieved a contact resistance of a few ohms.
This process created different microstructures allowing direct probing of in-plane and out-of-plane electrical transport, with different current orientations corresponding to the [$2\bar{1}\bar{1}0$], [$\bar{1}2\bar{1}0$], and [0001] directions
(hereafter labeled as $a$, $a'$, and $c$, respectively).
The microstructures are labeled as L1$a$, L1$a'$, L2$a'$, L2$c$, and L3$c$, indicating lamella Li (i = 1–3) combined with the current direction along $a$, $a'$, and $c$.
In Fig.~\ref{fig:a'-axis}, we show an example of a scanning electron microscopy (SEM) image of the microstructure L1$a'$.
The structure has a cross-section of $(4.3 \times 2.1)\,\mu$m$^2$ and a length between the voltage contacts of $32.1\,\mu$m. 

Multiple voltage leads are patterned along each bar to measure both longitudinal and transverse voltages.
The bar widths and contact separations are kept comparable between the different orientations so that any differences in magnetotransport can be attributed primarily to the crystallographic direction rather than to the geometry.

We measured the resistance using a standard AC four-point lock-in technique.
We performed measurements on microstructures in a commercial 14\,T magnet system.
In addition, we carried out pulsed-field measurements in a 70\,T pulse magnet (pulse duration of 150\,ms).
For the pulsed-field measurements we applied excitation currents of 100-\SI{500}{\micro\ampere} with a detection frequency of 30\,kHz. During each pulse, the raw AC voltages were recorded and subsequently analyzed using a digital lock-in and filter procedure.

For calculations of the electronic band structure of CrSb we applied two different density-functional-theory (DFT) software packages.
We found excellent agreement in the results from both approaches.

First, we performed DFT calculations of the band structures and Fermi surfaces using the projector-augmented wave (PAW) method~\cite{blochl1994,Kresse1999} implemented in the Vienna ab initio simulation package (\textsc{vasp})~\cite{Kresse1996cms, Kresse1996prb}. 
We used the experimentally determined crystal-structure parameters from Ref.~\cite{Yuan2020}.
The exchange-correlation functional was treated within the generalized gradient approximation (GGA) parameterized by Perdew, Burke, and Ernzerhof~\cite{Perdew1996}.
In the self-consistent calculations, the cutoff energy for the plane-wave expansion was 600 eV, and the $k$-point sampling grid of the Brillouin zone was $ 28\times28\times28 $, for both calculations with and without SOC.

Second, in order to check the reliability of the first results we performed DFT calculations utilizing another software package, the full-potential local-orbit minimum basis code (\textsc{fplo}, version 22.00-26)~\cite{Koepernik1999}.
We used the crystal-structure parameters from Ref.~\cite{Noda1983}.
To introduce the magnetic order, we removed all symmetry operations that generate the atom sites from the Wyckoff positions and introduced the additional Wyckoff positions $[0, 0, 1/2]$ for Cr and $[2/3, 1/3, 3/4]$ for Sb.
This maintains the crystal structure and allows the two Cr Wyckoff positions to have equal but antiparallel magnetic moments.  
We focused on calculations using the well-established local spin-density approximation (LSDA)~\cite{Perdew1992}. 
Additionally, we conducted exploratory \textsc{fplo} calculations with the GGA, the modified Becke-Johnson approximation (mBJ)~\cite{Tran2009}, shifted Fermi levels, as well as LDA$+$U and GGA+U with a potential applied to the Cr-$3d$ states.
None of these provided a clearly improved description of the experimental data.
We conducted scalar-relativistic calculations to visualize the altermagnetic behavior of CrSb and full-relativistic calculations to determine the bands, FSs, and quantum-oscillation frequencies, including SOC.
We used a mesh with $28\times28\times28$ points to calculate the density of states and band structure, and a finer grid with 112 points per dimension for calculating the FSs.

\section{Results and Discussion}

\begin{figure}[tb]
    \centering
    \includegraphics[width=\columnwidth]{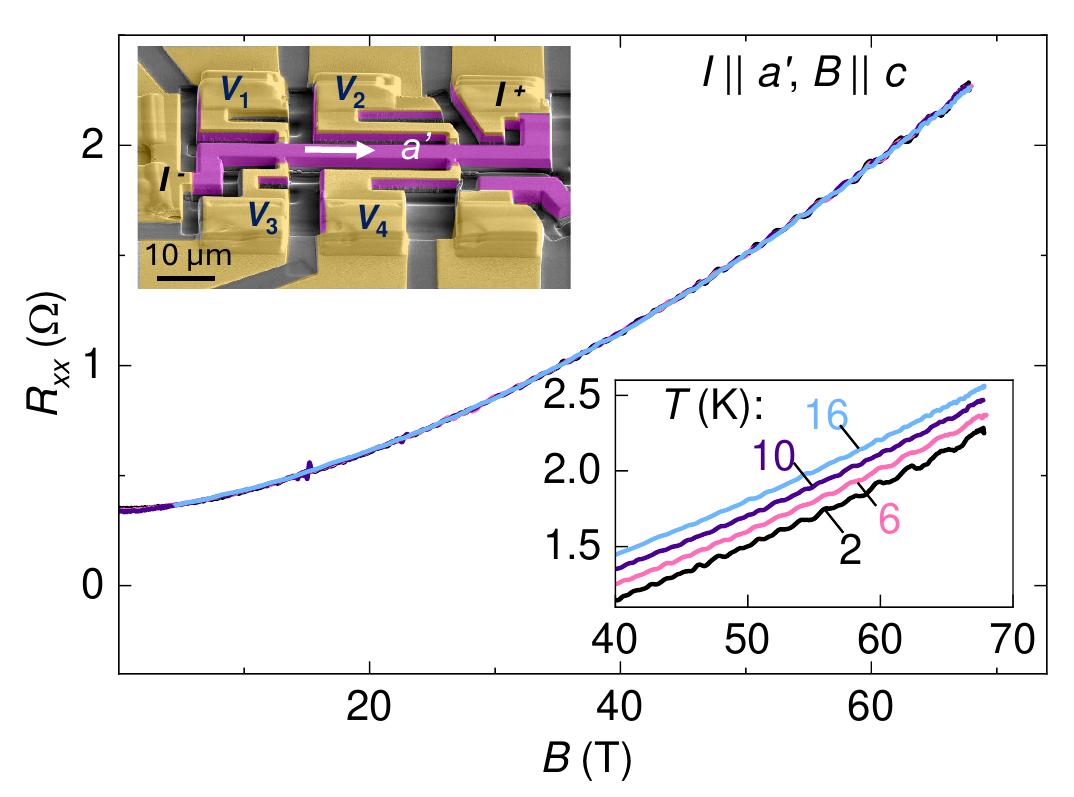}
    \caption{Field dependence of the longitudinal resistance of sample L1$a'$ up to 67\,T for selected temperatures. 
     Upper inset: False-color SEM image of the microstructure L1$a'$ with current applied along $a'$ axis and voltage measured between $V_1$ and $V_2$.
    Purple shaded regions indicate bulk CrSb, while yellow shaded regions represent sputtered gold contacts.
    Lower inset: Zoom-in with curves shifted by a constant offset of $0.1\,\Omega$.
    Slow variations are SdH oscillations.}
    \label{fig:a'-axis}

\end{figure}

We conducted magnetotransport measurements at fixed temperatures, $T$, in pulsed fields up to 67\,T for two samples, L1$a$ and L1$a'$, simultaneously with current running along the $a$ and $a'$ direction, respectively.
In Fig.~\ref{fig:a'-axis}, we present selected magnetoresistance curves recorded for L1$a'$ at $T=2$, 6, 10, and 16\,K for field applied along the $c$ direction.
Clear SdH oscillations, arising from the Landau quantization of well-defined FS orbits, are observable at lowest temperatures that subside upon increasing $T$.

We performed a basic characterization of the magnetoresistance and Hall effect in steady and pulsed fields.
In Fig.~\ref{fig:L1-MR}, we provide example curves recorded for L1$a$ and L1$a'$ at 4.2, 78, and  300\,K.
We observe a nonsaturating MR and a nonlinear Hall effect consistent with previous reports~\cite{Du2025,urata2024,bai2025,peng2025}, associating this with the multiband nature of the electronic band structure of CrSb.

\begin{figure}[tb]
    \centering
    \includegraphics[width=\columnwidth]{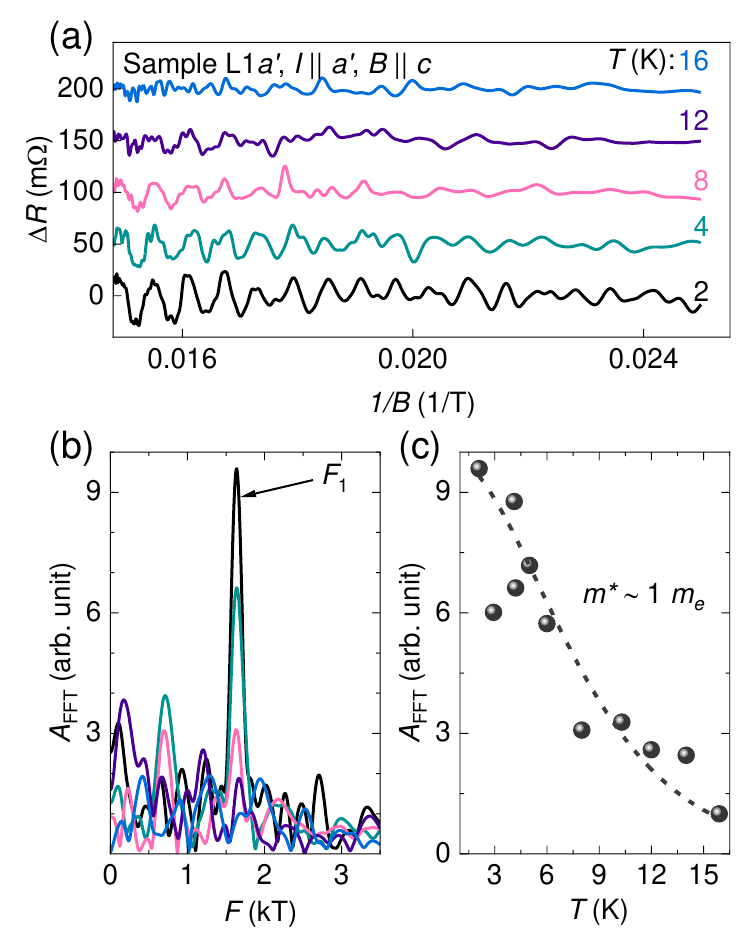}
    \caption{
    (a) Selected residual-resistance curves for L1$a'$ after subtraction of a third-degree polynomial for different temperatures plotted against inverse magnetic field aligned along $c$.
    (b) FFT spectra of the data in (a) for a field window between 40 and 66.6\,T.
    (c) Temperature dependence of the FFT amplitude for the frequency $F_1$.
    The dashed curve is a fit using the LK thermal damping factor.}
    \label{fig:a-T dependence}
\end{figure}

In Fig.~\ref{fig:a-T dependence}, we present the temperature dependence of the observed SdH oscillations in L1$a'$.
Figure~\ref{fig:a-T dependence}(a) shows the residual magnetoresistance $\Delta R$ after subtracting a monotonic background.
We subtracted a third-degree polynomial fit to the background and plot the relative oscillatory component versus the inverse magnetic field $1/B$.
The oscillation amplitude progressively decreases with increasing temperature.
The SdH oscillations are observable up to about 12\,K.
This confirms their origin in a single dominant extremal cross-section of the FS.
The periodicity in $1/B$ and the absence of beating indicate that this particular orbit is not affected by multi-frequency interference, making it suitable for a quantitative Lifshitz–Kosevich (LK) analysis~\cite{Shoenberg1984}.
We performed a fast-Fourier-transform (FFT) analysis of the oscillatory signals.
The FFT spectra for a field window between 40 and 66.6\,T of selected temperatures are plotted in Fig.~\ref{fig:a-T dependence}(b).
We observed a dominant sharp peak at approximately 1640\,T (labeled $F_1$).

\begin{figure}[tb]
    \centering
    \includegraphics[width=\columnwidth]{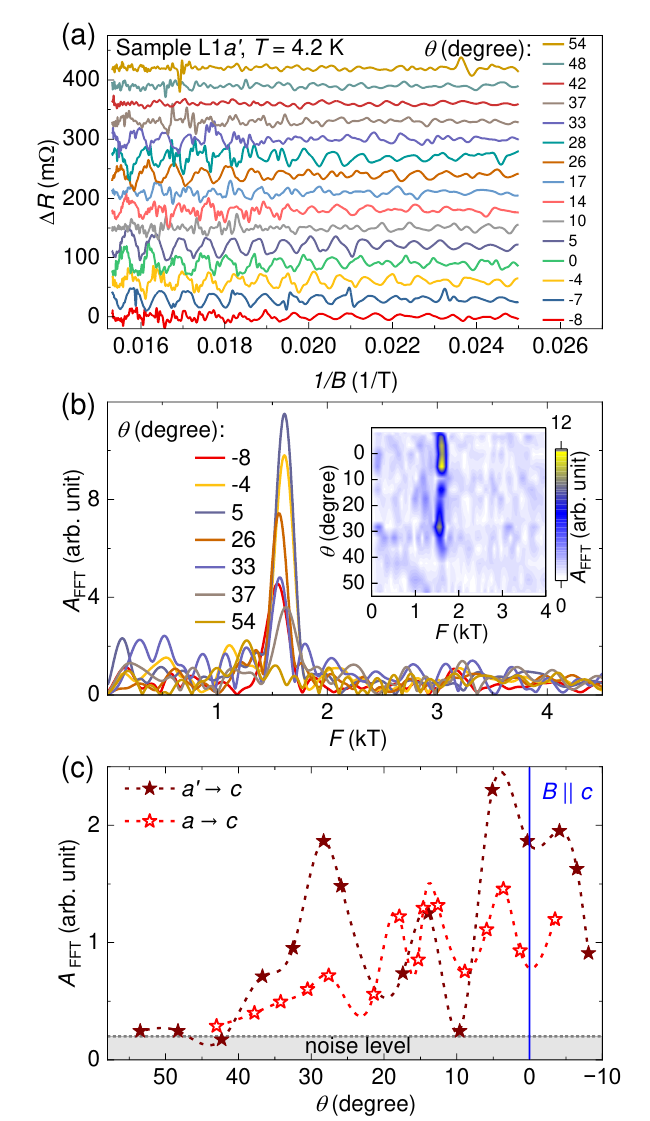}
    \caption{
    (a) Magnetoresistance of L1$a'$ recorded at 4.2\,K after background subtraction (third-degree polynomial) for various fixed polar angles $\theta$ within the $a'-c$ plane.
    (b) Respective FFT spectra for selected angles.  
    The inset shows a contour plot of the FFT amplitudes.
    (c) Angle-dependent FFT amplitude for the peak in the FFT spectrum for rotations in the $a'-c$ and $a-c$ plane.
    Dashed lines are guides to the eye using a basis spline in order to highlight the oscillating nature of the MQO amplitude depending on the tilt angle. These measurements are done separately and are analyzed for different field windows.}
    \label{fig:a'-theta}
\end{figure}
With the help of the Onsager relation $F = (\hbar/2\pi e\,) A_{\mathrm{F}}$, where $F$ is the frequency, $\hbar$ is the reduced Planck constant, and $e$ is the electron charge, we estimate the corresponding FS cross-sectional area $A_{\mathrm{F}} \approx 0.156\,\text{\AA}^{-2}$.
Figure~\ref{fig:a-T dependence}(c) shows the $T$ dependence of the FFT amplitude for $F_1$.
The dashed curve is a fit using the LK thermal damping factor, $R_\mathrm{T} =\alpha m^\ast T/[B\sinh{(\alpha m^\ast T / B)}]$, with $\alpha = -14.69$\,T/K, $m^\ast$ being the effective cyclotron mass, and $B$ is the harmonic mean of the field window.
This relation yields $m^\ast\approx 1m_\mathrm{e}$ for this orbit.

We further conducted magnetoresistance measurements on L1$a'$ at 4.2\,K in pulsed fields up to 67\,T for various polar tilt angles $\theta$, where $\theta=0^\circ$ denotes $B\parallel c$.
In Fig.~\ref{fig:a'-theta}(a), we present the background-subtracted (third-degree polynomial) $\Delta R$ plotted against the inverse magnetic field .
We present the respective FFT spectra in Fig.~\ref{fig:a'-theta}(b).
For large $\theta$, the oscillations are weak and close to the noise level. Still the amplitude of the oscillations grows significantly towards $c$ with some intermediate minima. This is better visible in the contour plot [inset of Fig.~\ref{fig:a'-theta}(b)], which shows the dependence of the FFT amplitude with respect to $F$ and $\theta$.

Figure~\ref{fig:a'-theta}(c) shows the angular dependence of the oscillation amplitude of $F_1$ for the rotation from $a'$ to $c$ (brown stars) as well as for a second angular study [L1$a'$(2)] for a rotation from $a$ to $c$ (red stars).
These oscillation-amplitude variations may originate from a complex three-dimensional shape of the underlying FS.
For a quasi-two-dimensional FS, the frequency is expected to follow a $1/\cos \theta$ increase.
However, $F_1$ remains nearly constant with angle, hinting at a spherical three-dimensional shape.

We additionally fabricated microstructures L2$a'$ and L2$c$ (Fig.~\ref{fig:L2-ZFR}) with the current applied along the $a'$ and $c$ direction, respectively.
Selected data are shown in Fig.~\ref{fig:L2-all}.
Interestingly, we could resolve SdH oscillations only below about $60^\circ$ for current applied along $a'$ (L2$a'$).
For current along $c$ (L2$c$), however, we could not detect any oscillations close to $B\parallel c$.
Only for larger angles, we resolve weak, low-frequency SdH oscillations below about 1\,kT [Figs.~\ref{fig:L2-all}(d) and (f)].
\begin{figure}[tb]
    \centering
    \includegraphics[width=\columnwidth]{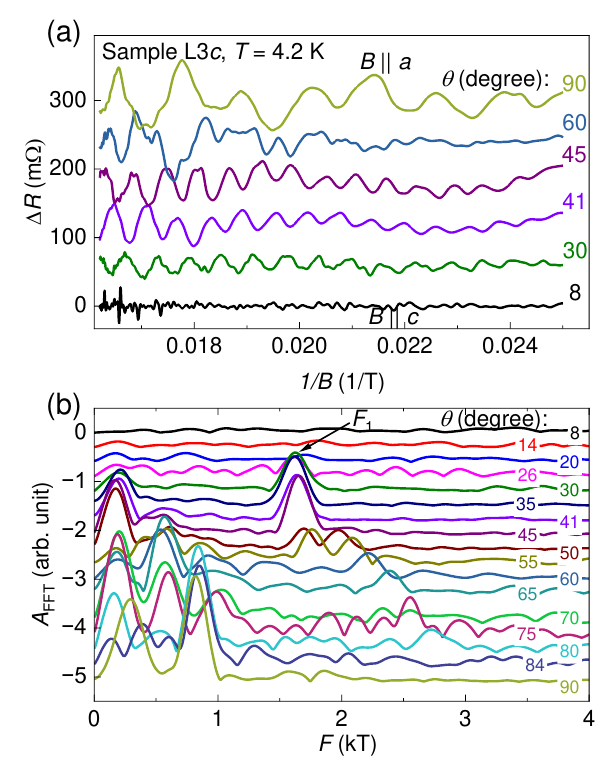}
    \caption{(a) Magnetoresistance of sample L3$c$ after background subtraction of a third-degree polynomial for selected polar angles $\theta$ within the $a-c$ plane.
    (b) FFT spectra shifted consecutively by a constant offset.
    We chose a field window between 40 and 61.7\,T.}
    \label{fig:L3c}
\end{figure}

Finally, we fabricated sample L3$c$, which exhibits a strongly enhanced residual resistance ratio (RRR) of about 47 (Fig.~\ref{fig:L3-ZFR}) reflecting an exceptional crystal quality.
We were able to resolve SdH oscillations with various frequencies from different bands over a broad angular range above $\theta \approx 26^\circ$ (Fig.~\ref{fig:L3c}).
As highlighted in Fig.~\ref{fig:L3c}(b), the SdH frequency $F_1 \approx 1640$\,T appears for angles between 26 and $45^\circ$.
At $\theta=50^\circ$, the FFT amplitude of $F_1$ drops significantly and its position shifts towards higher frequencies.
Despite the high sample quality, we observed no MQO frequencies exceeding 3\,kT, as were reported by another group from  magnetic torque measurements~\cite{Long2026}.
In addition, a second peak appears at about 2\,kT, as well shifting to higher frequencies as we increase $\theta$.
Starting at about $35^\circ$, clear low-frequency oscillations appear at about 250 and 600\,T.
The latter frequency increases slightly to about 800\,T with larger $\theta$.
This is similar to what we observed in L2$c$.
This collection of SdH frequencies provides a quantitative basis for comparison with band-structure calculations and for identifying the orbits responsible for the observed SdH signals.

\section{DFT calculations}
\begin{figure}[tb]
    \centering
    \includegraphics[width=\columnwidth]{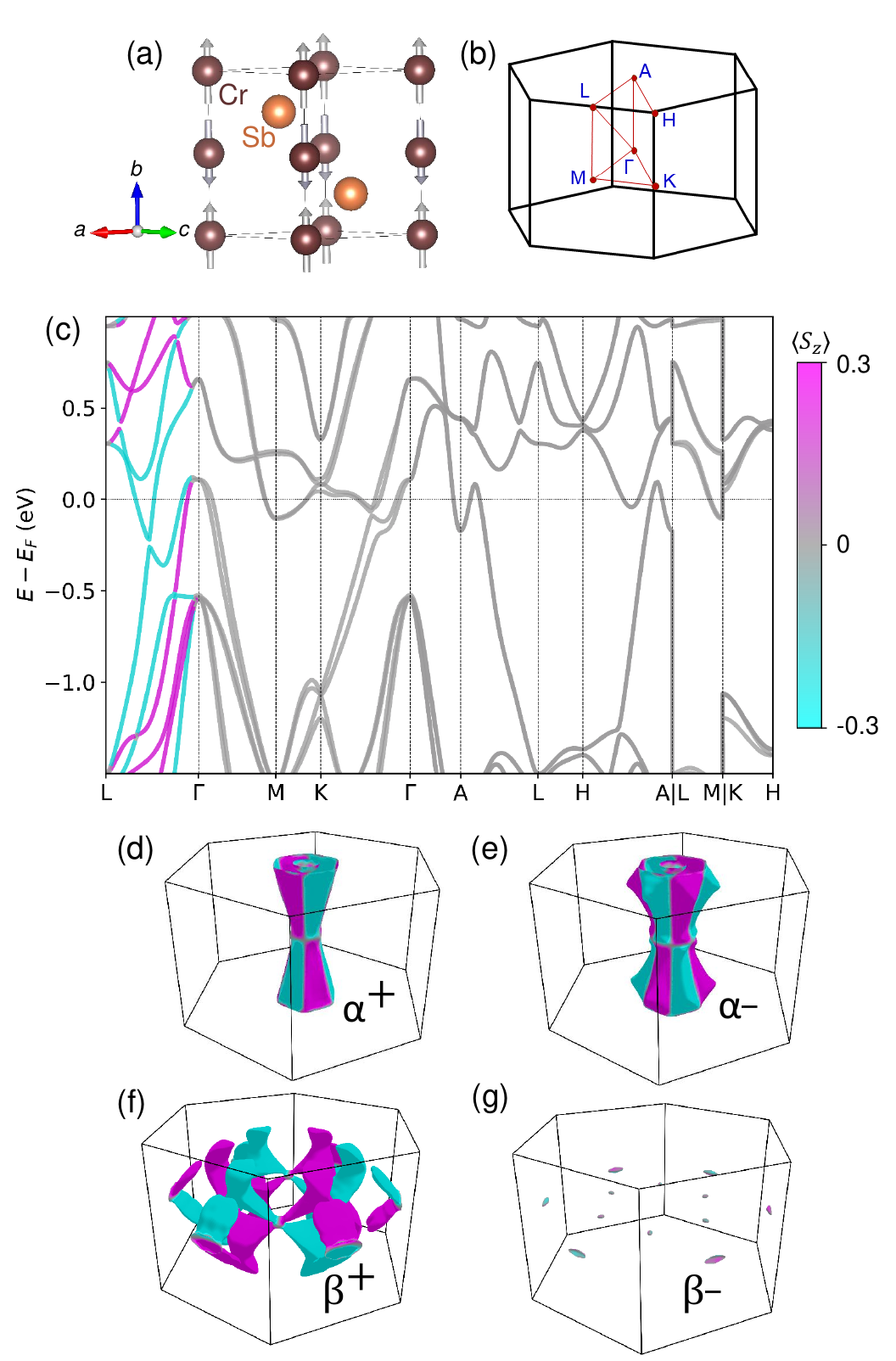}
    \caption{
    (a) Crystal structure of Cr Sb
    (b) Sketch of the first Brillouin zone and high-symmetry points.
    (c) Altermagnetic band structure of CrSb including SOC using \textsc{VASP}.
    (d)-(g) Calculated FSs in the first Brillouin zone, where $+$ and $-$ indicate the band shifts due to the magnetic exchange.
    Purple- and cyan-shaded areas mark the local spin polarization $S_z$ along the $c$ axis.
    }
    \label{fig:band structure}
\end{figure}
CrSb crystallizes in the hexagonal NiAs-type crystal structure [Fig.~\ref{fig:band structure}(a)], where the Cr atoms form hexagonal close-packed layers stacked along the $c$ axis.
The Sb atoms are located at octahedral interstitial sites between the Cr layers~\cite{Snow1952, Takei1963}. 
We performed DFT calculations using two different approaches, namely \textsc{VASP} and \textsc{fplo}, which both produced very similar results. 
Figure~\ref{fig:band structure}(b) shows the high-symmetry paths in the first Brillouin zone.

The band structure calculated for nonmagnetic CrSb can be found in Appendix~\ref{App:B} in Fig.~\ref{fig:FS-nonmagnetic}(a).
In the nonmagnetic case, the band structure of CrSb would consist of spin-degenerate bands, which stands in stark contrast to the spin splitting reported from experiments and AM models~\cite{Smejkal2022,Smejkal2022b, reimers2024, Ding2024, Shi2025, Du2025}. 
This acts as a baseline proving that altermagnetic order explicitly needs sublattices with opposite magnetic moments to capture the $k$-dependent splitting near the Fermi level, which is impossible to obtain in nonmagnetic states due to energy-band degeneracy.
Moreover, the predicted MQO frequencies from these nonmagnetic calculations [Fig.~\ref{fig:FS-nonmagnetic}(b)] differ significantly from the experimental SdH data, further confirming the necessity of AM calculations.

The Appendix Fig.~\ref{fig:BS-without SOC} shows the band-structure calculation of AM CrSb without SOC. 
Our experimental results agree best with the calculations that include SOC. 
Hence, we proceeded with LSDA and PAW calculations that include SOC for further analysis.
Figure~\ref{fig:band structure}(c) shows our results of (PAW+SOC) band-structure calculations with dispersion relations along high-symmetry paths in the Brillouin zone.
\begin{figure}[tb]
    \centering
    \includegraphics[width=\columnwidth]{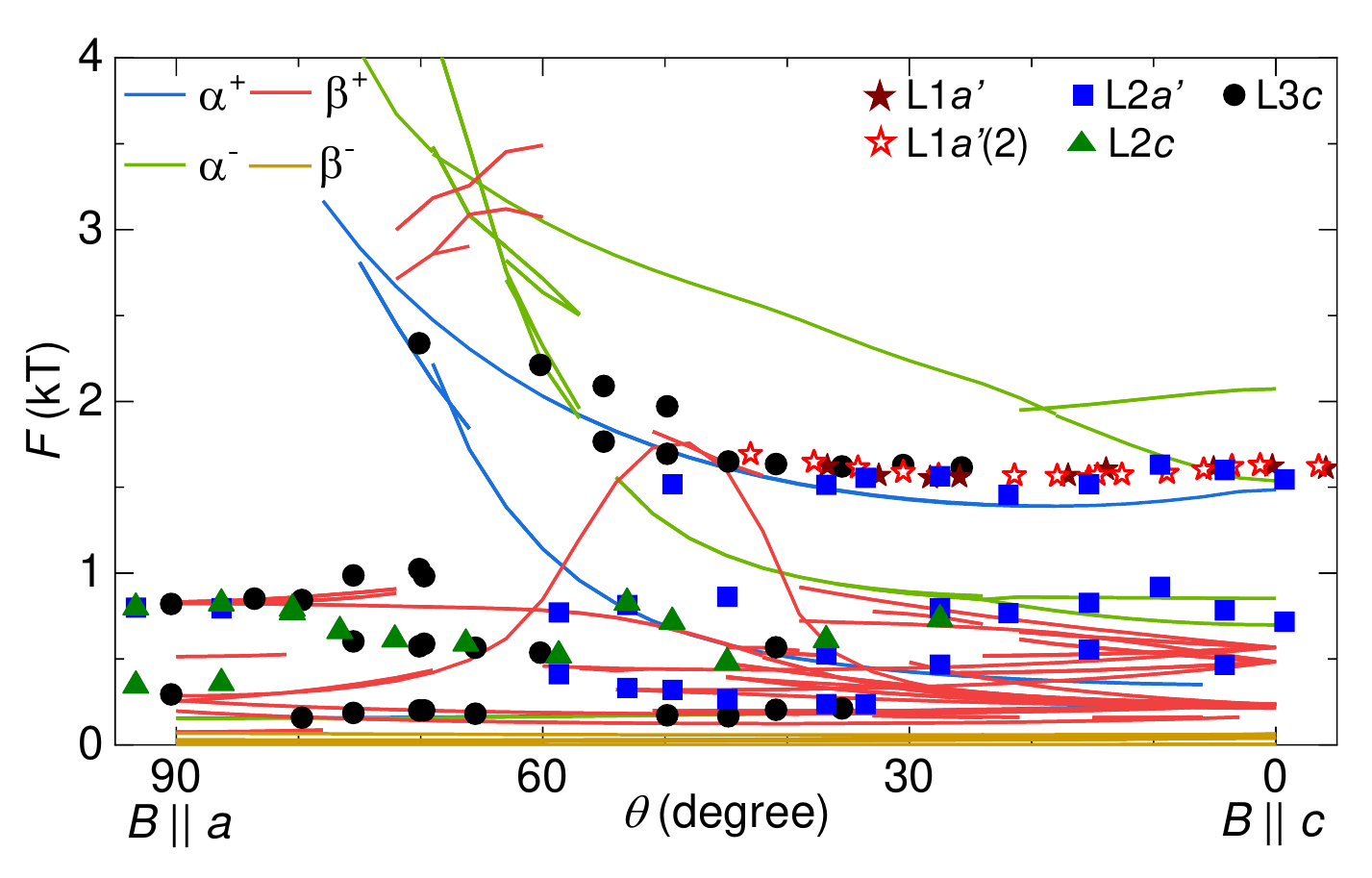}
    \caption{Angular dependence of experimentally observed (symbols) and calculated (lines) SdH frequencies.
    The experimental points represent the most prominent maxima in the FFT spectra for rotations from $c$ to $a'$ and from $c$ to $a$, respectively, for samples L1 (pink and orange), L2 (blue and green), and L3 (black).
    The lines correspond to those of the calculated extremal orbits shown in Figs.~\ref{fig:band structure}(d) - \ref{fig:band structure}(g).
    The SdH frequencies are calculated using \textsc{fplo} with LSDA and SOC.
    }
    \label{fig:f-theta}
\end{figure}

While spin splitting along the $\Gamma$-M-K path is driven by spin-orbit coupling and remains below 0.2\,eV, the $\Gamma$-L path explicitly exhibits AM spin splitting, which can reach values approaching the 1\,eV scale.
The results reveal two spin-split bands ($\alpha^{+}$, $\alpha^{-}$, $\beta^{+}$, $\beta^{-}$) crossing the Fermi level, where $+$ and $-$ indicate the band shifts due to the magnetic exchange.
From these calculations, we obtain the FSs shown in Figs.~\ref{fig:band structure}(d) - \ref{fig:band structure}(g), where purple- and cyan-shaded areas mark the local spin polarization along the $c$ axis.  
They are comprised of highly anisotropic hole-like spin-split $\alpha$ bands, a complicated multi-connected electron-like $\beta^{+}$ band, and multiple small electron-like pockets for the $\beta^{-}$ band. 
These features are important for reconciling the observed MQOs.

\begin{table}[tb]
  \centering
  \caption{
  Calculated SdH frequencies ($F$) and  band masses ($m_b$) for $B \parallel c$.
  The subscripts (1 to a maximum of 5) denote different extremal areas of the respective FS.
  }
  \begin{tabular}{lcc}
    \hline
    Orbit & $F$ (T) & $m_b$ ($m_e$) \\
    \hline
    $\beta^-_1$        &  42  & 0.69 \\
    $\beta^-_2$        &  63  & 1.51 \\
    $\beta^+_1$          & 226  & 1.65 \\
    $\alpha^+_1,\alpha^-_1$ & 226 & 0.34 \\
    $\beta^+_2$          & 236  & 0.33 \\
    $\beta^+_3$          & 485  & 0.65 \\
    $\beta^+_4$          & 568  & 1.16 \\
    $\alpha^-_2$       & 699  & 0.47 \\
    $\alpha^-_3$       & 853  & 0.88 \\
    $\alpha^+_2$         & 1485 & 0.78 \\
    $\alpha^-_4$       & 1537 & 0.79 \\
    $\alpha^-_5$       & 2073 & 0.92 \\
  \hline
  \end{tabular}
  \label{tab:freq_mass}
\end{table}

Figure~\ref{fig:f-theta} compares the angular dependence of the measured SdH frequencies against the calculated extremal orbits (taking SOC into account).
The colored lines represent the calculated frequencies for the shown distinct Fermi‑surface sheets.
The experimental data appear as colored symbols, where each color denotes frequencies obtained from one sample.
Table~\ref{tab:freq_mass} shows the calculated extremal orbits together with the corresponding cyclotron band masses $m_b$.

We find an overall agreement between the experimental data and calculated frequencies.
For instance, the high‑frequency of the $\alpha^{+}_2$ branch around $F\approx 1485$\,T fits nicely to the measured SdH frequency $F_1$ and its angular dependence observed in samples L1$a'$, L2$a'$, and L3$c$; the SdH frequencies observed in L3$c$ between $\theta=30^\circ$ and $B\parallel a$ match well with calculated orbits of the $\beta^{+}$ band.

Note: Two recent preprints report the observation of MQOs as well~\cite{Long2026, Terashima2026}. 
Our results match closely those reported from electrical resistance measurements recorded on bulk crystals in steady fields by Terashima \textit{et al.}~\cite{Terashima2026}.
In contrast, the work by Long \textit{et al.}~\cite{Long2026} reports high-frequency de Haas-van Alphen oscillations (dHvA) detected via magnetic torque on bulk crystals, which they associate with a "dog-bone"-shaped three-dimensional Fermi surface.
SdH and dHvA oscillation amplitudes scale differently, which may explain why certain orbits may or may not be discernible depending on the detection method~\cite{Shoenberg1984}. 
In  order to make their DFT results match, Long \textit{et al.} applied significant shifting of individual bands by up to 100\,meV.
Our work yields a good agreement between the DFT results and experimental data without the need for shifting bands.


The experimentally determined effective mass ($m^* = 1m_\mathrm{e}$) for $F_1$ [Fig.~\ref{fig:a-T dependence}(c)] is somewhat larger than the calculated value ($m_b = 0.78m_e$), which is reasonable since in the calculated value no mass enhancements due to many-body interactions are included. 
This validates that the DFT result reproduces the FS of CrSb reliably.
Only some frequencies around 500\,T observed in the samples L2$c$ and L3$c$ between $\theta=70^\circ$ to $76^\circ$, cannot be assigned to calculated orbits. 
Further work is needed to verify the experimental data and to optimize the DFT calculations.

\section{Conclusion}

In summary, we have successfully grown high-quality CrSb single crystals and carried out detailed magnetotransport studies up to 68\,T.
Our observations reveal a nonsaturating magnetoresistance and pronounced Shubnikov-de Haas oscillations.
In addition, we unravel a complex, anisotropic multiband electronic structure in CrSb, which is composed of spin-split $\alpha$ and $\beta$ bands with cyclotron masses of the order of $1m_e$.
Angle-dependent Shubnikov-de Haas data observed on multiple samples closely track the SdH frequencies computed from density-functional-theory calculations.
The comparison of our experimental results to these calculations confirms the altermagnetic nature of CrSb's electronic band structure, as it is essential in order to capture the observed spin splitting that does not arise in the nonmagnetic state.
Our findings not only validate the predicted altermagnetic Fermi-surface topology of CrSb, but also highlight the essential role of extreme magnetic fields in resolving its multiband electronic landscape.

\section{Acknowledgments}

We would like to acknowledge the support by S. Findeisen in the design and fabrication of pulse-field probes.
Access to the Ion Beam Center (IBC) of the HZDR is gratefully acknowledged.
We acknowledge support from the Deutsche Forschungsgemeinschaft (DFG) through Grant No. HE 8556/3-1 and the
W\"{u}rzburg-Dresden Cluster of Excellence on Complexity, Topology and Dynamics in Quantum Matter--$ctd.qmat$ (EXC 2147, Project No.\ 390858490), as well as the support of the HLD at HZDR, member of the European Magnetic Field Laboratory (EMFL).
We acknowledge funding from the Czech Science Foundation (Grant No. 22-22000M), and Lumina Quaeruntur fellowship LQ100102201 of the Czech Academy of Sciences.
Part of this work was funded by the European Union as part of the Horizon Europe call HORIZON-INFRA-2021-SERV-01 under grant agreement number 101058414 and co-funded by UK Research and Innovation (UKRI) under the UK government’s Horizon Europe funding guarantee (grant number 10039728) and by the Swiss State Secretariat for Education, Research and Innovation (SERI) under contract number 22.00187.
Views and opinions expressed are however those of the author(s) only and do not necessarily reflect those of the European Union or the UK Science and Technology Facilities Council or the Swiss State Secretariat for Education, Research and Innovation (SERI). Neither the European Union nor the granting authorities can be held responsible for them.

\appendix

\section{Magnetotransport measurements}
\label{App:A}
\setcounter{figure}{0}\renewcommand{\thefigure}{A\arabic{figure}}
\begin{figure}[b]
    \centering
    \includegraphics[width=0.9\columnwidth]{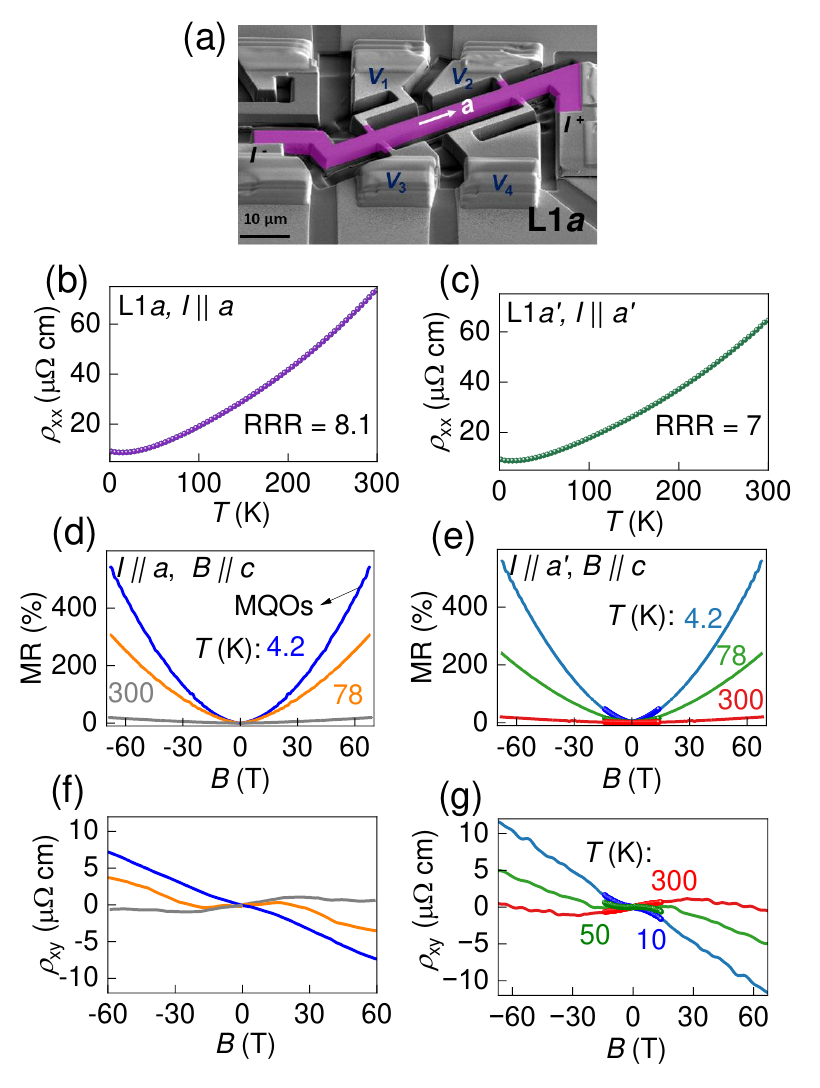}
    \caption{
    (a) False-color SEM image of the FIB-structured sample L1$a$. 
    (b, c) Temperature dependence of the zero-field resistivity, (d, e) the magnetoresistance, and (f, g) the Hall resistivity of the microstructures L1$a$ and L1$a'$, respectively.
    For L1$a'$, the thick lines show data taken in static fields up to 14\,T.}
    \label{fig:L1-MR}
\end{figure}
In  Fig.~\ref{fig:L1-MR}(a), we show an SEM image of the microstructure L1$a$.
From the temperature dependence of the resistivity, we observe RRRs of 8 for L1$a$ [Fig.~\ref{fig:L1-MR}(b)] and 7 for L1$a'$ [Fig.~\ref{fig:L1-MR}(c)].
For the analysis, we symmetrized the measured longitudinal resistivity ($\rho_{xx}$) and anti-symmetrized the Hall resistivity data ($\rho_{xy}$) with respect to the external applied magnetic field according to 
\begin{equation}
\rho_{xx}(B) = \frac{\rho_{xx}(+B) + \rho_{xx}(-B)}{2},
\label{symmetrisation}
\end{equation}
\begin{equation}
\rho_{xy}(B) = \frac{\rho_{xy}(+B) - \rho_{xy}(-B)}{2}.
\label{antisymmetrisation}
\end{equation}
With this approach, we effectively eliminate artifacts introduced from geometric irregularities in the sample shape or the positioning of the contacts.
The room-temperature resistivity of 65 and  $73\,\mu\Omega\mathrm{cm}$ for the microstructures L1$a$ and L1$a'$, respectively, are consistent with recent reports that range from $57\,\mu\Omega\mathrm{cm}$~\cite{bai2025} to $70\,\mu\Omega\mathrm{cm}$~\cite{peng2025}.

\begin{figure}[tb]
    \centering
    \includegraphics[width=0.9\columnwidth]{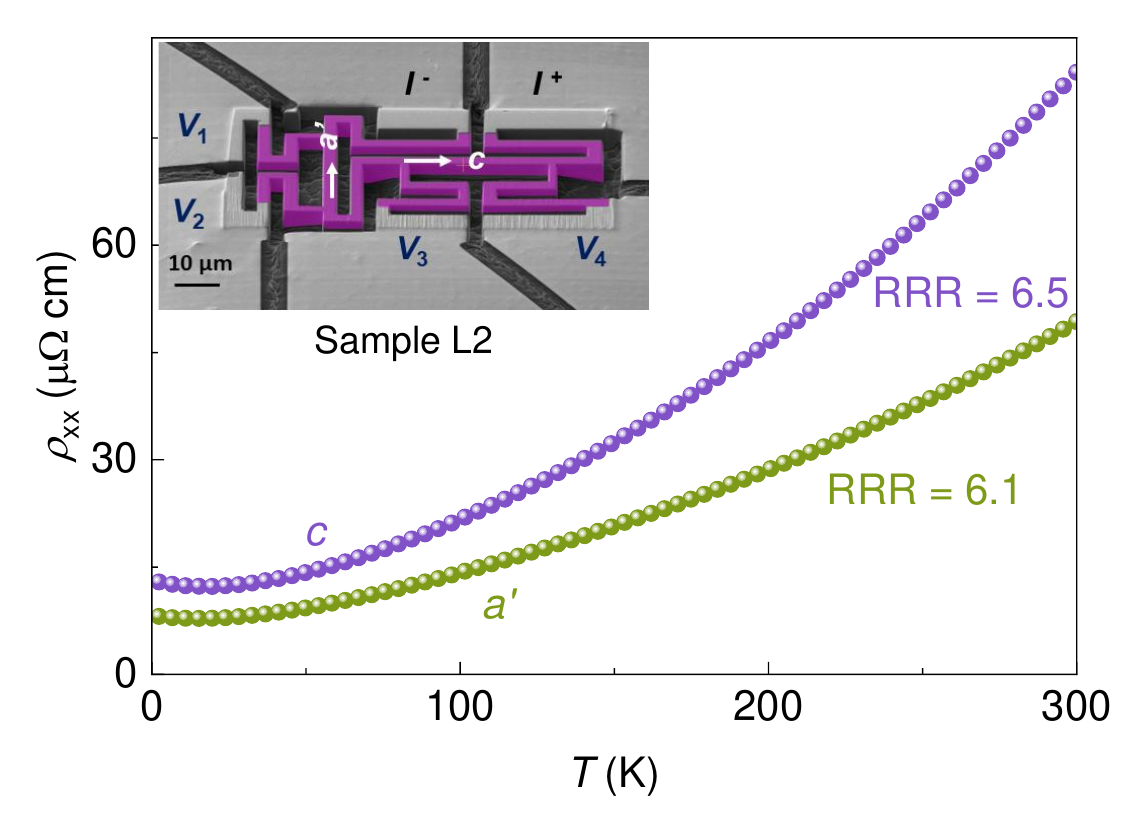}
    \caption{
    Temperature dependence of the zero-field resistivity using the voltage drops between $V_1$ and $V_2$ ($I\parallel a'$) and $V_3$ and $V_4$ ($I\parallel c$).
    Inset: False-color SEM image of the microstructure L2 that allows to probe $c$ (horizontal) and $a'$-axis (vertical) transport simultaneously.
    }
    \label{fig:L2-ZFR}
\end{figure}
The magnetoresistance, MR $= \rho_{xx}(B)/\rho_0 -1$, exhibits a nearly quadratic, nonsaturating response up to 65\,T. 
For L1$a$ [Fig.~\ref{fig:L1-MR}(d)], the MR at 4.2\,K reaches $543\%$, while at 300\,K it is suppressed to $19\%$ at 65\,T. 
For L1$a'$ [Fig.~\ref{fig:L1-MR}(e)], the MR reaches $559\%$ at 4.2\,K and $20\%$ at 300\,K temperature.
Careful inspection of both MR curves reveals subtle differences in the curvature and absolute values at the highest fields, indicating anisotropic Fermi velocities along the two crystallographic directions. 
The MRs show clear SdH oscillations in both structures at higher magnetic fields. 

The Hall resistivities $\rho_{xy}$ [Figs.~\ref{fig:L1-MR}(f) and \ref{fig:L1-MR}(g)] show nonlinear field dependencies across all temperatures. 
This nonlinearity indicates the coexistence of multiple electron and hole bands with varying mobilities, as evidenced by our band-structure calculations.
The thick lines in Figs.~\ref{fig:L1-MR}(e) and \ref{fig:L1-MR}(g) show data up to 14\,T recorded during static-field measurements.
These data are in very good agreement with the pulse-field measurements. 

The inset in Fig.~\ref{fig:L2-ZFR} shows a false-color SEM image of microstructure L2 with current paths along the $c$ and $a'$ axes.
The $c$-axis structure (resistance measured between $V_3$ and $V_4$) has a cross-section of $(2.6 \times 3.5)\,\mu$m$^2$ and a length of $34.1\,\mu$m, whereas the $a'$-axis structure (resistance measured between $V_1$ and $V_2$) has a cross-section of $(2.7 \times 3.6)\,\mu$m$^2$ and a length of $25.4\,\mu$m. 
Figure~\ref{fig:L2-ZFR} shows the corresponding zero-field resistances of each structure.

\begin{figure}[H]
    \centering
    \includegraphics[width=\columnwidth]{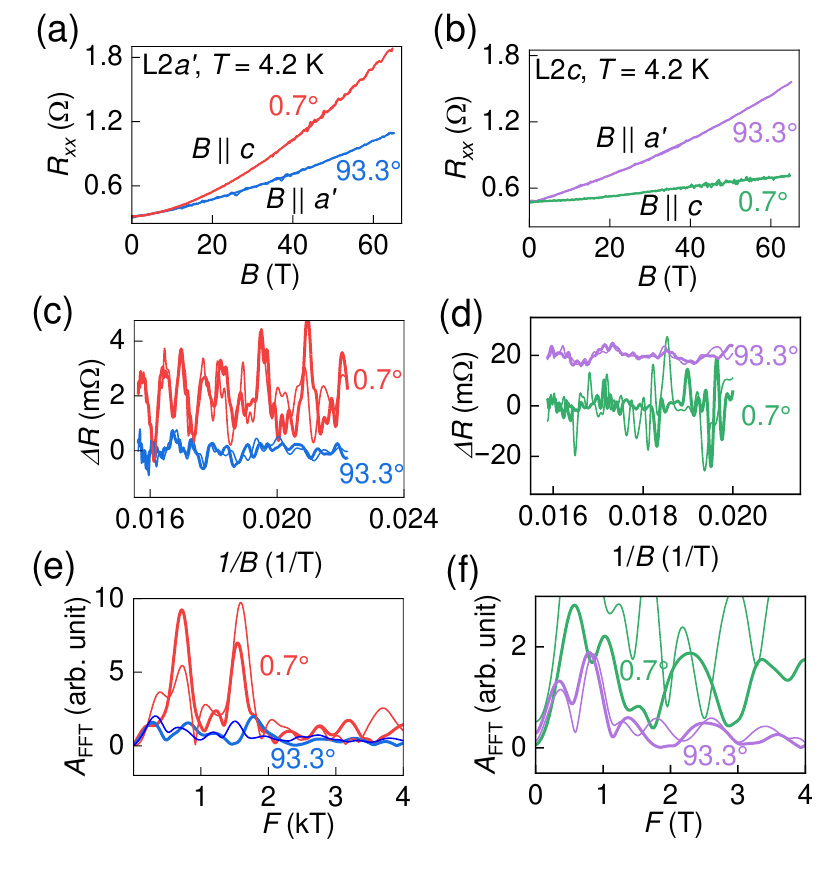}
    \caption{
    Longitudinal magnetoresistance recorded in pulsed field up to 62\,T for microstructure L2 with  current applied along the $a'$ (a) and $c$ axis (b),respectively, for two  field orientations. 
    (c, d) Background-subtracted resistance (third-degree polynomial fit) for $I\parallel a'$ (c) and $I\parallel c$ (d), respectively. 
    Corresponding FFT spectra from field windows 45–64\,T (e) and 50–64\,T (f). 
    Bold (thin) lines indicate up (down) field sweeps.
    }
    \label{fig:L2-all}
\end{figure}
\begin{figure}[tb]
    \centering
    \includegraphics[width=1\columnwidth]{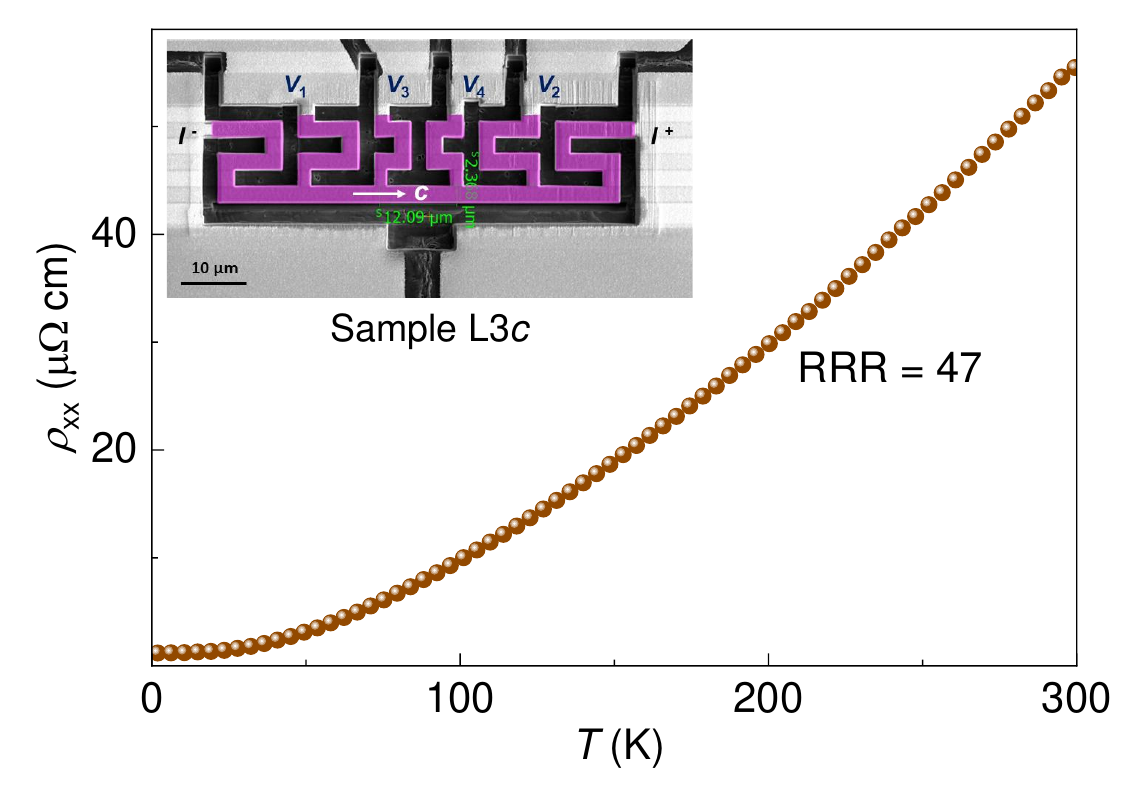}
    \caption{
    Temperature dependence of the zero-field resistivity of L3$c$.
    Inset: False-color SEM image of the microstructure. }
    \label{fig:L3-ZFR}
\end{figure}

Figures~\ref{fig:L2-all}(a) and \ref{fig:L2-all}(b) show the longitudinal magnetoresistances of L2$a'$ and L2$c$, respectively.
Figure~\ref{fig:L2-all}(c) and \ref{fig:L2-all}(d) show the residual resistance $\Delta R$ after subtracting a monotonic background.
While Figs.~\ref{fig:L2-all}(e) and \ref{fig:L2-all}(f) show the corresponding amplitude of the SdH oscillations from the FFTs.
The data exhibit pronounced anisotropies in the MR.
At 62\,T the MR for $B \parallel c$ reaches about $500\%$ in L2$a'$ [Fig.~\ref{fig:L2-all}(a)], but only $50\%$ for in L2$c$ [Fig.~\ref{fig:L2-all}(b)].

The inset of Fig.~\ref{fig:L3-ZFR} shows a false-color SEM image of microstructure L3$c$, which has a cross-section of $(2.4 \times 2.2)\,\mu\mathrm{m}^2$ and a length of $36\,\mathrm{\mu m}$ between contacts $V_1$ and $V_2$.
The enhanced RRR of 47 indicates high crystalline quality and a long electronic mean free path. 
We confirmed the elemental composition of L3$c$ using energy-dispersive x-ray spectroscopy, yielding a 1:1 ratio of Cr to Sb. 
This rules out significant crystal impurities or the presence of foreign elements in the lattice. 
\begin{figure}[tb]
    \centering
    \includegraphics[width=\columnwidth]{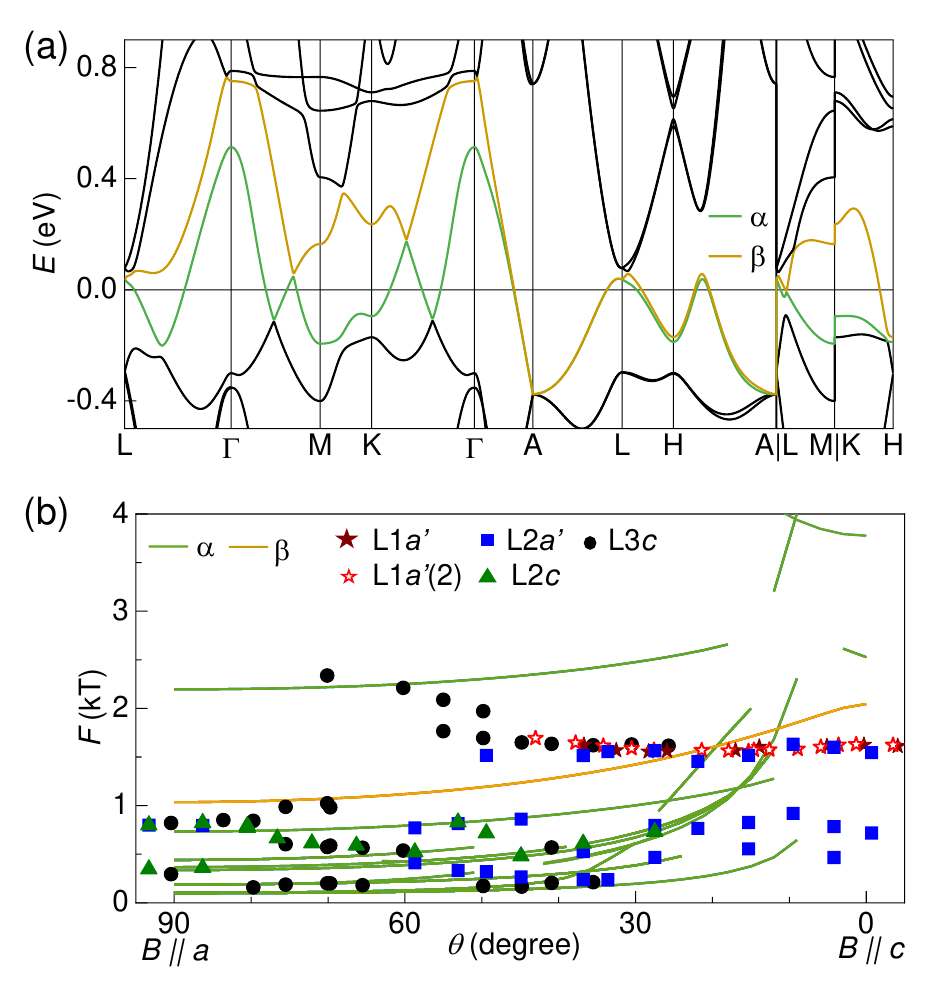}
    \caption{
    (a) Electronic band structure along high-symmetry paths in the nonmagnetic state calculated using \textsc{fplo} with LSDA.
    (b) Comparison of the angular dependence of experimentally observed (symbols) and calculated (lines) SdH frequencies for the nonmagnetic case.
    }
    \label{fig:FS-nonmagnetic}
\end{figure}

\section{Band-structure calculations}
\label{App:B}

Figure \ref{fig:FS-nonmagnetic}(a) shows the calculated electronic band structure of CrSb for the nonmagnetic state using \textsc{fplo} with LSDA. 
The absence of significant spin splitting is expected for calculations of a nonmagnetic structure.  
Figure \ref{fig:FS-nonmagnetic}(b) shows the angular dependence of the calculated MQO frequencies (solid lines), which strongly deviate from experimental SdH data (symbols).

\begin{figure}[tb]
    \centering
    \includegraphics[width=\columnwidth]{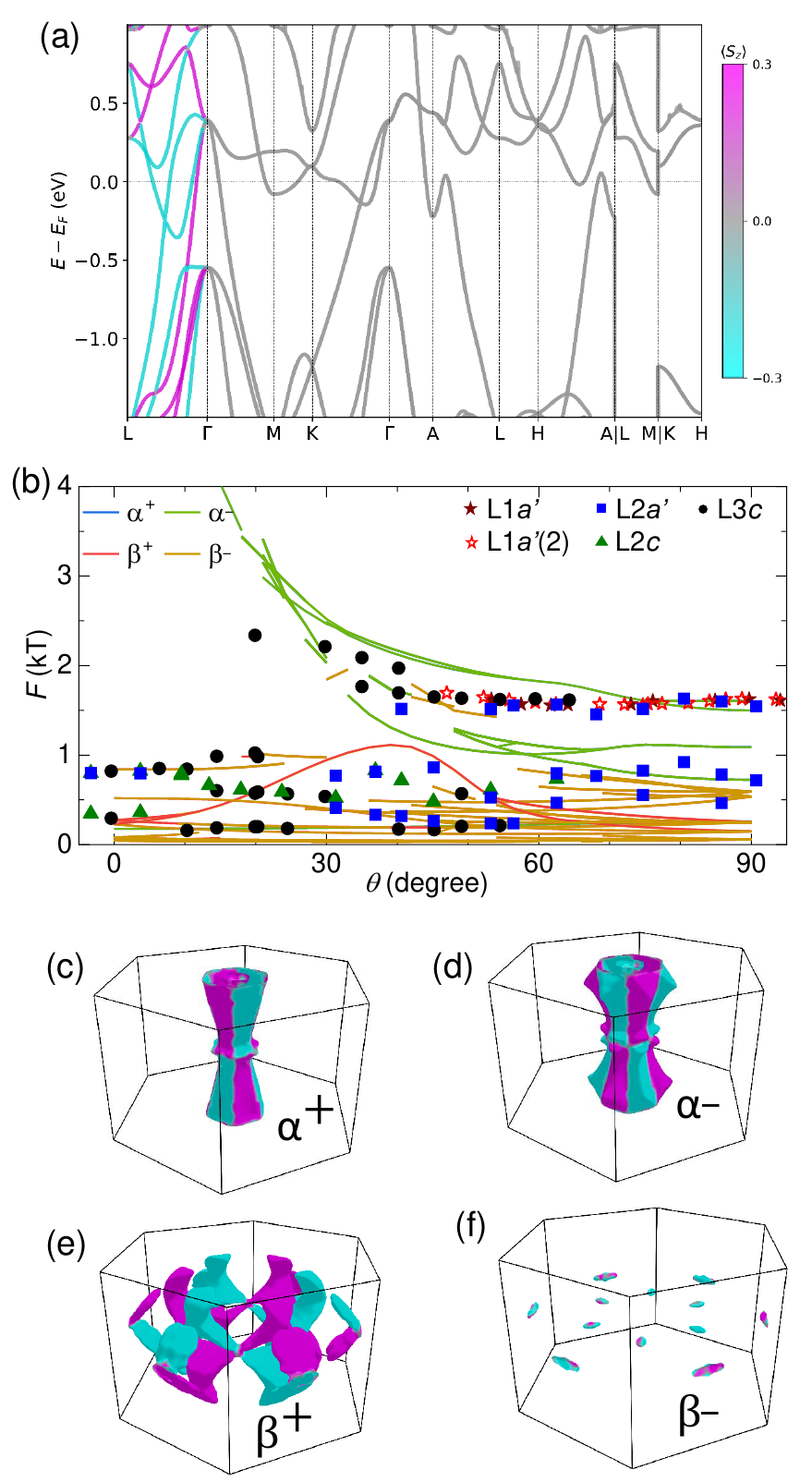}
    \caption{(a) Electronic band structure along high-symmetry paths in the altermagnetic magnetic state without SOC using the PAW method in \textsc{vasp}.
    (b) Comparison of the angular dependence of experimentally observed (symbols) and calculated (lines) SdH frequencies for the AM case without SOC using \textsc{fplo}.
    (c)-(f) Calculated FSs in the first Brillouin zone.}
    \label{fig:BS-without SOC}
\end{figure}
Figure~\ref{fig:BS-without SOC}(a) shows the calculated band structure of AM CrSb without spin-orbit coupling. 
Some paths such as $\Gamma$-L, reveal characteristic altermagnetic spin splittings near the Fermi level.
The results reveal two spin-split bands (\(\alpha^{+}\), \(\alpha^{-}\), \(\beta^{+}\), \(\beta^{-}\)) crossing the Fermi level, similar to the band-structure calculation with SOC (Fig.~\ref{fig:band structure}).
In Fig.~\ref{fig:BS-without SOC} (b), we compare the measured SdH frequencies (symbols) with calculated values (lines), which reflects significant discrepancies.
In particular, the experimentally determined angular dependence of $F_1$ is not reproduced by the calculations. 
In comparison, the band structure including SOC fits much better with the experimental data (Fig.~\ref{fig:f-theta}), which is why we chose the LSDA+SOC model to describe our data.

\begin{figure}[tb]
    \centering
    \includegraphics[width=\columnwidth]{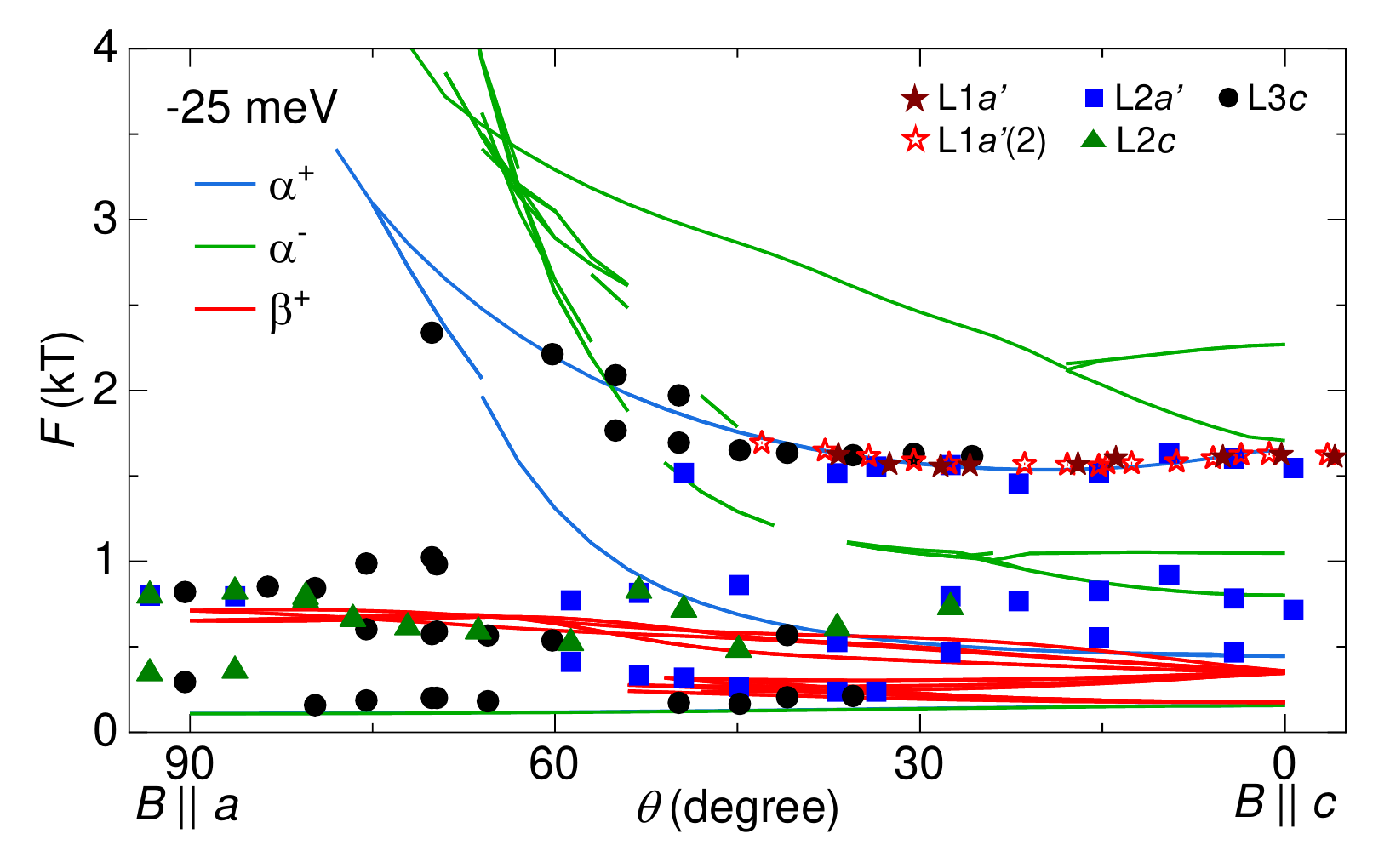}
    \caption{Comparison of the angular dependence of experimentally observed (symbols) and calculated (lines) SdH frequencies for the AM case, including SOC, using \textsc{fplo} with \textsc{LSDA} and an artificial shift of the Fermi level by -25\,meV.
    }
    \label{fig:f-theta-25meV}
\end{figure}

We also tried to adjust the Fermi level artificially in order to obtain a better match between the experimental data and the calculated ones.
In Fig.~\ref{fig:f-theta-25meV}, we applied a shift of -25\,meV to the Fermi level.
With that, the $\alpha^+$ branch around 2\,kT now fits better to the experimental data.
However, the $\beta^-$ branch vanishes completely and the calculated low-frequency branches of $\beta^+$ describe the experimental data much less accurately.

\clearpage
\bibliography{Bibliography}

\end{document}